\documentstyle[twoside,art10]{article}
\catcode`@=11
\def\section{\@startsection {section}{1}{\z@}{-3.5ex plus-1ex minus
    -.2ex}{2.3ex plus.2ex}{\reset@font\center\bf}}
\def\thebibliography#1{\section*{\refname\@mkboth
  {\uppercase{\refname}}{\uppercase{\refname}}}\list
  {\@biblabel{\arabic{enumiv}}}{\settowidth\labelwidth{\@biblabel{#1}}%
    \usecounter{enumiv}%
    \let\p@enumiv\@empty
    \def\theenumiv{\arabic{enumiv}}}%
    \def\newblock{\hskip .11em plus.33em minus.07em}%
    \sloppy\clubpenalty4000\widowpenalty4000
    \sfcode`\.=1000\relax}
\def\ps@myheadings{\let\@mkboth\@gobbletwo
 \def\@oddhead{\hfil{\sl\rightmark}\hfil}%
 \def\@oddfoot{\hfil\rm\thepage}\def\@evenhead{\hfil{\sl\leftmark}\hfil}%
 \def\@evenfoot{\rm\thepage\hfil}\def\sectionmark##1{}
\def\subsectionmark##1{}}
\@addtoreset{equation}{section}

\catcode`@=12
\def\be{\begin{equation}}
\def\ee{\end{equation}}

\def\q{{\exp(2\pi i/3)}}
\def\qq{{\exp(-2\pi i/3)}}
\def\limit{{q\to\q}}
\def\ff{f(\theta)}
\def\drr{{d\over d\theta}}

\def\pt{{\partial\over\partial\theta}}
\def\dt{{\partial\over \partial t}}

\def\dz{\partial_t}
\def\qm{\lim_{q\to\q}}

\def\ss{{\sum_{m=0}^\infty}}
\def\d{{\cal D}}
\def\ie{{\it i.e.}}
\catcode`@=11
\newdimen\z@ \z@=0pt
\def\m@th{\mathsurround=\z@}
\def\ialign{\everycr{}\tabskip\z@skip\halign} 
\def\eqalign#1{\null\,\vcenter{\openup\jot\m@th
  \ialign{\strut\hfil$\displaystyle{##}$&$\displaystyle{{}##}$\hfil
      \crcr#1\crcr}}\,}
\def\matrix#1{\null\,\vcenter{\normalbaselines\m@th
    \ialign{\hfil$##$\hfil&&\quad\hfil$##$\hfil\crcr
      \mathstrut\crcr\noalign{\kern-\baselineskip}
      #1\crcr\mathstrut\crcr\noalign{\kern-\baselineskip}}}\,}
\catcode`@=12
\begin{document}
\begin{center}
{\bf BRAIDED STRUCTURE OF FRACTIONAL $Z_3$-SUPERSYMMETRY}
\footnote{Presented at $5^{th}$ Colloquium `Quantum groups and integrable 
systems', Prague, 20-22 June 1996.}
\\
\vspace{0.5cm}
\setcounter{footnote}{3}
{\sc 
J. A. de Azc\'arraga$^\dagger$, 
R. S. Dunne$^\ddagger$, 
\\
A. J. Macfarlane$^\ddagger$ 
and J. C. P\'erez Bueno$^\dagger$
\footnote{E-mails: 
azcarrag@evalvx.ific.uv.es, r.s.dunne@damtp.cam.ac.uk, \\
$\phantom{AAAAAAAi}$
a.j.macfarlane@damtp.cam.ac.uk and pbueno@lie.ific.uv.es.}
}
\begin{small}
\\[2mm]
{\sl $\dagger$ Departamento de F\'{\i}sica Te\'orica and IFIC, Centro Mixto 
Universidad de Valencia-CSIC,}
{\sl E-46100 Burjassot (Valencia) Spain.}
\\[2mm]
{\sl $\ddagger$ Department of Applied Mathematics \& Theoretical Physics,}
\\
{\sl University of Cambridge, Cambridge CB3 9EW, U.K.}
\end{small}
\end{center}

\vspace{0.5cm}
\noindent
{\small
It is shown that fractional $Z_3$-superspace is isomorphic to the 
$\limit$ limit of the braided line. 
$Z_3$-supersymmetry is identified as translational invariance along this line. 
The fractional translation generator and its associated covariant derivative
emerge as the $\limit$ limits of the left and right derivatives  
from the calculus on the braided line.
}
                                                    
\section{Brackets and q-grading}
\hskip\parindent
Our aim here is to reformulate some results of a previous paper \cite{AM}, 
where the structure of fractional supersymmetry was investigated from a group 
theoretical point of view, from a braided 
Hopf algebra approach. We shall not be concerned here with the possible 
applications of fractional supersymmetry and will refer instead to 
\cite{AM,DMPA} for references on this aspect. 

We begin by defining the bracket
\be
[A,B]_{q^r}:=AB-q^r BA
\quad,\label{two}
\ee
where $q$ and $r$ are just arbitrary complex numbers. 
If we assign an integer grading $g(X)$ to each element $X$ of some algebra, 
such that $g(1)$=0 and
$ g(XY)=g(X)+g(Y),$ 
for any $X$ and $Y$, we can define a bilinear graded $z$-bracket as follows,
\be 
[A,B]_z=AB-q^{-g(A)g(B)}BA\quad,\quad z=q^{-g(A)g(B)}\quad.
\label{four}
\ee
Here $A$ and $B$ are elements of the algebra, and of pure grade. 
The definition may be extended to mixed grade terms using the 
bilinearity. 
We also have
\be 
[r]_q:={{1-q^r}\over{1-q}}\quad,\quad
[r]_q!:=[r]_q[r-1]_q[r-2]_q...[2]_q[1]_q\quad,
\label{six}
\ee
supplemented by $[0]_q!=1$. 
When $q$ is $n$-root of unity the previous grading scheme becomes degenerate,
so that in effect the grading of an element is only defined modulo $n$. 
In this case also have $[r]_q=0$ when $r$ modulo $n$ is zero $(r\neq0)$. 

\section{$q$-calculus and the braided line}
\hskip\parindent
Consider the braided line \cite{MajI,MajII}, a simple deformation of the 
ordinary line characterized by a single parameter $q$.
Our braided line Hopf algebra will consist of a single variable $\theta$, 
of grade 1, upon which no additional conditions are placed
for generic $q$, by which we mean that $q$ is not a root of unity. 

The braided Hopf structure of this deformed line is as follows.
It has braided a coproduct,
\be
\Delta\theta=\theta\otimes1+1\otimes\theta\quad,
\label{seven}
\ee
\be
(1\otimes\theta)(\theta\otimes1)=q\theta\otimes\theta\quad,
\quad
(\theta\otimes1)(1\otimes\theta)=\theta\otimes\theta\quad,
\label{nine}
\ee
the braiding being given by
${\cal B}(\theta\otimes\theta)=q\theta\otimes\theta$.
There are also a counit and antipode,
\be
\varepsilon(\theta)=0\quad,\quad
S(\theta^r)=q^{r(r-1)\over2}(-\theta)^r\quad,
\label{oneone}
\ee
which satisfy all the usual Hopf algebraic relations,
as long as the braiding is remembered. From 
the braided Hopf algebra perspective, the coproduct generates a shift 
along the braided line. 
To bring this out more clearly we use a group-like notation for the coproduct
and write 
\be
\theta=1\otimes\theta\quad,\quad
\delta\theta=\epsilon=\theta\otimes1\quad,
\label{onetwo}
\ee
so that (\ref{nine}) leads to
\be
[\epsilon,\theta]_{q^{-1}}=0\quad.
\label{onethree}
\ee
In this form, the coproduct (\ref{seven}) expresses the additive group law,
\be
\Delta\theta=\epsilon+\theta \quad
\label{onefour}
\ee
which on the fractional Grassmann variable $\theta$ corresponds \cite{AM} 
to the action of the left translation $L_\epsilon$, 
$L_\epsilon\theta\equiv\theta'=\epsilon+\theta$.

The above expressions provide a basis upon which to construct 
a differential 
(and integral \cite{DMPA}) calculus on the braided line. 
We can introduce an algebraic left differentiation operator $\d_L$, 
in analogy with the undeformed case, 
via the requirement
$[\epsilon \d_L,\theta]=\epsilon,$ 
which implies that 
\be
[\d_L,\theta]_q=1\quad,\quad
(\drr\theta=1)\quad.
\label{onesix}
\ee  
This corresponds to defining [cf. (\ref{four})] the left derivative $\d_L$ by
\be
[\d_L,\theta]_z:=1\quad,\quad
z=q\quad.
\label{onesixa}
\ee
Regarding (\ref{onefour}) as the definition the 
left translation by $\epsilon$, and identifying 
$\d\equiv\d_L$, we can go on 
considering right shifts $R_\eta:\theta\mapsto\theta+\eta$ 
of parameter $\eta$ where 
$[\theta,\eta]_{q^{-1}}=[\eta,\theta]_q=0$. 
Reasonings similar to the above lead us to a right derivative operator 
$\d_R$,
which satisfies 
\be
[\theta,\d_R]_z=[\theta,\d_R]_q:=1\quad,\quad
([\d_R,\theta]_{q^{-1}}=-q^{-1})\quad.
\label{oneseven}
\ee
It may be shown that the left and right derivative operators are in general 
related \cite{DMPA} by
\be
\d_R=-q^{-(1+N)}\d_L\quad,
\label{oneeight}
\ee 
where $N$ is a number-like operator satisfying,
\be
[N,\theta]=\theta\quad,\quad 
[N,\d_L]=-\d_L\quad,
\label{onenine}
\ee
and consequently $[N,\d_R]=-\d_R$. 
This implies that $ [\d_L,\d_R]_{q^{-1}}=0$ or, alternatively, 
$[\epsilon\d_L,\eta\d_R]=0$ (commutation of the left $L_\epsilon$ and right 
$R_\eta$ shifts).

Let $f(\theta)$ be a function of $\theta$ defined by the positive power 
series expansion,
\be 
f(\theta)=\ss {C_m\theta^m\over[m]_q!}\quad,
\label{twoone}
\ee
where the $C_m$ are ordinary numbers. The derivative of $f(\theta)$ is 
generated by the graded bracket (\ref{onesix}) as follows,
\be
\eqalign{\drr \ff&:=[\d,\ff]_z
=\ss C_m\left[\d,{\theta^m\over[m]_q!}\right]_z
=\ss C_m\left[\d,{\theta^m\over[m]_q!}\right]_{q^m}\cr
&=\ss C_m{\theta^{m-1}\over[m-1]_q!}\quad.\cr}
\label{twotwo}
\ee
This clearly reduces to 
the calculus on the undeformed line when $q=1$. 
The left translation \ie, the coproduct (\ref{onefour}) is given 
\cite{DMPA,AM} by a deformed exponentiation (see 
\cite{GR,MajIV,McAnally}) of $\epsilon\d_L$.
We shall refer to the differential
calculus defined by (\ref{onesix}), (\ref{oneseven}) and 
(\ref{oneeight}) as $q$-calculus.

\section{$q$-calculus in the $q$ root of unity limit and $(Z_3)$ fractional
supersymmetry}
\hskip\parindent
When $q^m=1$, $[m]_q=0\ (m\ne 1)$ 
and expressions such as ${\theta^m\over [m]_q!}$
can be made meaningful only by assuming that $\theta^m$ is also zero.
In that case, we may identify the limit 
$\lim_{q\to\exp(2\pi i/m)}{\theta^m\over [m]_q!}$
with a degree zero (`bosonic') variable $t$.
It was shown in \cite{PLB} that for the $m=2$ case this procedure leads to a 
braided interpretation of supersymmetry,
the $Z_2$-graded group structure of which was discussed in \cite{AldAz}.
We shall consider here in detail the case of $Z_3$ fractional supersymmetry, 
the group analysis of which may be found in \cite{AM} (the general $Z_n$ case 
will be discussed in \cite{DMPA}).
The limit relevant here is the $\limit$ limit.
To take this limit we note that for $q$ not a root of unity 
we have in general the relationships
\be
\left[\d_L,{\theta^m\over[m]_q!}\right]_{z}=
\left[\d_L,{\theta^m\over[m]_q!}\right]_{q^m}=
{\theta^{m-1}\over[m-1]_q!}=
\left[{\theta^m\over[m]_q!},\d_R\right]_{q^m}=
\left[{\theta^m\over[m]_q!},\d_R\right]_{z}\quad.
\label{twofive}
\ee
In taking the $\limit$ limit of the above formulae
we encounter difficulties when $m=3$ since $[3]_q=0$. 
But it is possible to retain (\ref{twofive}) 
by requiring that the  $\limit$ limit of 
$\theta^3\over[3]_q!$  be finite and nonzero. 
This in turn requires
$\theta^3\to 0$ as $\limit$. 
This is preserved under the left shift $\theta\to\epsilon+\theta$, 
since $(\epsilon+\theta)^3=0$ follows from (\ref{onethree}) 
when $\limit$ provided that $\theta^3=0=\epsilon^3.$ 
We now note that under complex conjugation
we have $\overline{[3]_q!}=q^{-3}[3]_q!$ (along the circle of radius 1). 
Then, in the $\limit$ limit the $q$-factorial $[3]_\q!$ is real. 
As a result, we define 
\be
t:=-\qm{\theta^3\over[3]_q!}\quad,
\label{twoseven}
\ee
where the - sign is introduced to compare easily with \cite{AM}.
Since $\theta$ is assumed real, $t$ will also be real. 
By using the identities
\be
\qm\left({[3r]_q\over[3]_q}\right)=\qm
\left({1-q^{3r}\over1-q^3}\right)=r
\quad,
\label{twoeight}
\ee 
\be
\qm\left({[3r+1]_q\over[1]_q}\right)=\qm\left({1-q^{3r+1}\over1-q}
\right)=1\quad,
\label{twonine}
\ee
and
\be
\qm\left({[3r+2]_q\over[2]_q}\right)=\qm\left({1-q^{3r+2}\over1-q^2}
\right)=1\quad,
\label{twoninea}
\ee
we have, for $p=0,1,2,$ 
\be
\qm\left({\theta^{3r+p}\over[3r+p]_q!}\right)=
{\theta^p\over[p]_\q!}{(-t)^r\over r!}
\quad.
\label{thirty}
\ee

As mentioned, the finite limit in (\ref{twoseven}) 
denoted by $t$ was introduced in order to make possible the $\limit$ limit of 
(\ref{twofive}) at $m=3$. 
Similar problems arise for all $m\geq 3$ \cite{DMPA},
and the importance of (\ref{thirty}) is that it shows that these can also 
be handled in terms of $t$.
Thus, any function $f(\theta)$ on the braided line at generic $q$ leads in the 
$\limit$ limit to a function of the form $f(t,\theta)$ 
(or `fractional superfield' on fractional superspace $(t,\theta)$). 
To investigate further the properties of $t$, and to see how it fits into our
$q$-calculus, let us now consider
\be
\left[\d_L,
\left[\d_L,
\left[\d_L,{\theta^3\over[3]_q!}\right]_{q^3}
\right]_{q^2}
\right]_q=1=
\left[
\left[
\left[
{\theta^3\over[3]_q!},\d_R
\right]_{q^3},
\d_R\right]_{q^2},\d_R\right]_q
\quad,
\label{threeone}
\ee
(notice the appropriate $q$-factor in each bracket depending on the 
grading of its components, cf. (\ref{twofive}), (\ref{four}))
valid for all $q\neq\q$. 
Taking the $\limit$ limit we see that (\ref{threeone}) 
reduces to $[\d_L^3,t]=-1=[t,\d_R^3],$
so that by identifying
\be
\dz=-\d_L^3=\d_R^3\quad,
\label{threethree}
\ee
we have
\be
[\dz,t]=1\quad,\label{threefour}
\ee
which is just the defining relation of the algebra associated with ordinary 
calculus. 

Let us consider the {\it left calculus}.
Using $\dz$ given by (\ref{threethree})
to induce differentiation with respect to $t$, the full $q$ calculus for 
$q=\q$ 
obtained from (\ref{twotwo}) and (\ref{twoseven}) is given by,
\be
\eqalign{        
        &\qquad\qquad\qquad\qquad \drr\theta 
        =[\d,\theta]_\q=1
        \quad,\cr
        \drr t&=[\d_L,t]=\lim_{\limit}[\d_L,-{\theta^3\over[3]_q!}]_{q^3}
        =-{\theta^2\over[2]_\q}=\q\theta^2\quad,\cr
        \dt t
        &
        =[\dz,t]=1\quad,\quad \dt\theta=[\dz,\theta]=-[\d_L^3,\theta]=
        -[\d_L,[\d_L,[\d_L,\theta]_q]_{q^0}]_{q^{-1}}=0\quad.\cr}
\label{threefive}
\ee
Since $\dt\theta=0$ and $\drr t\neq0$, we can only avoid a contradiction 
by interpreting $\dt$ as a partial derivative, and $\drr$ as a total 
derivative, a result which we implicitly took into account when
choosing our notation. 
We can also define partial differentiation with respect to $\theta$. We do
this as follows
\be
\pt\theta:=[\partial_\theta,\theta]_\q=1\quad,\quad
\pt t:=[\partial_\theta,t]=0\quad.
\label{threesix}
\ee
Using this definition we are able to perform a chain rule expansion of the 
total derivative, so that
\be
\drr={d\theta\over d\theta}\pt+{d t\over d\theta}\dt=\pt
-{\theta^2\over[2]_\q}\dt=\pt+\q\theta^2\dt
\quad.
\label{threeseven}
\ee
By substituting (\ref{threeseven}) into the definition (\ref{threethree}) 
we obtain an additional but expected condition,
\be
{\partial^3\over\partial\theta^3}=0\quad.
\label{threeeight}
\ee
This can all be put into the algebraic form $\d_L=\partial_\theta+
\q{\theta^2}\dz,$ 
where
\be
[\partial_\theta,\theta]_\q=1
\quad ([\theta,\partial_\theta]_\qq=-\qq)
\label{leftder}
\ee
and
$\partial_\theta^3=0.$

The {\it right calculus} may be introduced similarly.
Besides $\partial_t=\d_R^3$, we have
\be
\eqalign{
	{d_R\over d_R\theta}\theta=[\theta,\d_R]_\q=1\;,
	&\;
	{d_R\over d_R\theta}t=
        -\lim_\limit[{\theta^3\over [3]_q!},\d_R]_{q^3}=
	\exp(2\pi i/3)\theta^2\,,
	\cr
	{\partial\over\partial t}t=[\partial_t,t]=1\quad,
	&\quad
	{\partial\over\partial t}\theta=[\d_R^3,\theta]=0\quad.
        \cr}
\label{new}
\ee
Introducing a partial right derivative $\delta_\theta$ by
\be
[\theta,\delta_\theta]_\q:=1\quad ([\delta_\theta,\theta]_{\qq}=-\qq)\quad,
\label{rightder}
\ee
the expression of $\d_R$ differs by a sign from that of $\d\equiv\d_L$, 
namely
\be
{d_R\over d_R\theta}={\delta\over\delta\theta}+{\theta^2\over[2]_\q}\dt=
{\delta\over\delta\theta}-\q\theta^2\dt\ ,\ 
\d_R=\delta_\theta-\q{\theta^2}\dz\,.
\label{chainrule}
\ee
The $\delta_\theta$ introduced above differs from that found in \cite{AM} in a 
$-\q$ factor,
$\delta_\theta$(here)$=-\exp(-2\pi i/3)\delta_\theta(\hbox{ref.\ }\cite{AM})$

\section{$Z_3$-fractional supersymmetry from a braided point of view}
\hskip\parindent
If follows from the above that
$\d_L\equiv Q$ and $-\q\d_R\equiv D$ 
are just the supercharge and the corresponding covariant derivative 
encountered in $(Z_3)$ fractional supersymmetry \cite{AM}. 
Hence
\be
Q^3=-\dz\quad,\quad
D^3=-\dz\quad.
\label{algebra}
\ee
We may identify $Q$ and $D$ as,
respectively, the generators of left and right shifts along the braided 
line at $q=\q$.
These were shown in \cite{AM} to correspond to 
the right- $[Q]$ and left-invariant $[D]$ `fractional translation' 
generators. 

To further investigate this point of view we examine the Hopf structure on the 
braided line in the $\limit$ 
limit. When $q=\q$, (\ref{nine}) reduces to
\be
\eqalign{(1\otimes\theta)(\theta\otimes1)&=\q\theta\otimes\theta\quad,\quad
(\theta\otimes1)(1\otimes\theta)=\theta\otimes\theta\quad,\cr}
\label{fourseven}
\ee
so that from (\ref{seven}), we find
\be
\Delta\theta^3=\theta^3\otimes1+1\otimes\theta^3+
(1+\q+[\q]^2)(\theta\otimes\theta^2 + \theta^2\otimes\theta)=0\quad,
\label{fournine}
\ee
as required by the homomorphism property of the coproduct.
The counit and antipode take the following form
\be
\varepsilon(\theta)=0\quad,\quad
S(\theta)=-\theta\quad.
\label{fifty}
\ee

The braided structure (\ref{seven}),(\ref{fourseven}) is the 
standard one, see for example \cite{MajIII,MajII} for a
discussion of super and anyonic quantum groups. 
The new structure appears when the variable $t$ 
defined by the limit (\ref{twoseven}) is introduced \cite{PLB,DMPA}. From 
the definition (\ref{twoseven}) it follows that $\theta$ and $t$ commute. From 
(\ref{twoseven}) and (\ref{seven})-(\ref{nine}) 
we compute
$\qm\Delta [{-\theta^3\over[3]_q!}]$
using that $[n]_q=1+q+\ldots+q^{n-1}$ for generic $q$.
This, together with $[t,\theta]=0$ and (\ref{seven})-(\ref{oneone}), 
shows that
the algebra generated by $(t,\theta)$ has a braided Hopf algebra  
structure, with
\be
\Delta t=t\otimes1+1\otimes t+\q(\theta\otimes\theta^2 + \theta^2\otimes\theta)
\quad,\quad
\epsilon(t)=0\quad,\quad
S(t)=-t\quad,
\label{fiveone}
\ee
(for instance $\Delta([t,\theta])=0=[\Delta t,\Delta\theta]$).
This means that although $t$ and $\dz$ 
satisfy the algebra associated with ordinary (undeformed) calculus, $t$ has 
non primitive coproduct: 
the coaddition no longer corresponds to a time translation.
Considered along with the chain rule 
expansion of the $q$-calculus
derivative (\ref{threeseven}), it is clear that in the $\limit$ limit we 
cannot decompose the $q$-calculus
algebra into unrelated $t$ and $\theta$ parts. 
Indeed, from (\ref{fiveone}) 
we see that 
when the braided group is considered, the appearance of 
$\theta$ in the coproduct of $t$ means that no such decomposition can be 
performed. 
The fact that we cannot regard this braided Hopf algebra
as a product entity is an essential feature of fractional supersymmetry in 
general. 
To see this for the present $Z_3$-case we rewrite the coproducts of 
$\theta$ and $t$ using the notation (\ref{onetwo}).
Using (\ref{seven}), (\ref{fournine}) and definition (\ref{twofive}) 
we obtain
\be
\eqalign{\theta\to\Delta\theta=\epsilon+\theta\quad,\quad
t\to\Delta t=t+\tau+q(\epsilon\theta^2+\epsilon^2\theta)\quad,\cr}
\label{fivetwo} 
\ee
where now $q\equiv\q$ (so that $-1/[2]_q=q$) and
\be
\tau=\lim_{q\to\q}=-{\epsilon^3\over [3]_q!}
\label{fivethree}
\ee
is a time translation independent of $t$.
This is just the form 
of the finite $Z_3$-supersymmetry transformation of \cite{AM};
the transformation of $t$ follows from that of $\theta$ via the 
relationship (\ref{twoseven}) and the coproduct $\Delta$.

\section{Final remarks}
\hskip\parindent
To conclude, let us summarize our results \cite{PLB,DMPA} and
outline the new point of view which they provide. 
For generic $q$ the braided line described in section 3 is well defined. 
In the $\limit$ limit the nilpotency of  
$\theta$ prevents us from having a complete description of the braided line 
and its associated differential calculus. 
A convenient way in which we can obtain
such a description is to introduce an additional variable $t$,
defined as in (\ref{twoseven}). From 
(\ref{thirty}) this is seen to carry a structure which, 
for generic $q$, is related
to $\theta^3$ and higher powers of $\theta$. So in
the $\limit$ limit the braided line is made up
of the two variables $\theta$ and $t$, which span the one-dimensional 
fractional superspace.
Furthermore, under a shift along this braided line $\theta$ and $t$ transform 
exactly as in $Z_3$ fractional supersymmetry. 
Thus we are able to identify $Z_3$-superspace with the $q=\q$ limit of the 
braided line, and $Z_3$-supersymmetry as translational invariance along this 
line. 
Clearly, (fractional)
superspace cannot be regarded as the tensor product of independent 
$t$ and $\theta$ parts; it is instead a single braided geometric 
entity. 
This provides a
braided interpretation of the central extension character of the 
$Z_3$-graded group aspect of fractional supersymmetry discussed in \cite{AM}.
To conclude, we wish to stress that the above results are not restricted to 
the $Z_3$ case.
Similar results also hold for supersymmetry 
\cite{PLB} (cf. \cite{AldAz}) and in the $Z_n$ case \cite{DMPA}.

\section*{Acknowledgements}
This paper describes research supported in part by E.P.S.R.C and P.P.A.R.C.
(UK) and by the C.I.C.Y.T (Spain).
J.C.P.B. wishes to acknowledge an FPI grant
from the CSIC and the Spanish Ministry of Education and Science.

\end{document}